# Decelerated expansion or Future time comes to a halt

Charles B. Leffert

Available Energy, Inc.
1302 Wrenwood Dr.
Troy, Michigan 48084-2688

**ABSTRACT**
A new speculative model for the expansion of our universe has been under development by the author for the last two decades, which correctly predicts astronomical measurements with no dark matter or dark energy. This new closed model (no free parameters) correctly predicts that time increases without limit into the future. By contrast, a seldom mentioned future-problem in relativity theory is shown in this paper. Any acceleration of the expansion rate of our universe destroys the proper behavior of time into the future. The goal of this paper is to present this problem of relativity theory and remind the reader of the success of the new model

**Key words:** cosmology: theory; cosmology: cosmological parameters; time: general relativity:

## 1. INTRODUCTION
Over short periods of time and small distances, general relativity (GR) theory makes excellent predictions of the trajectory of particles inside our universe. However over long periods of time and large distances of the expansion of our universe, relativity theory is even worse than useless indeed, its predictions are misleading. There are many fundamental problems in the use of the big bang (BB) theory to account for the expansion of our universe.

The most egregious is the GR-predicted collapse of the closed universe. Nature may be harsh but it is not stupid. Not one form of life deliberately evolves toward its own demise, and neither does the universe itself.

Relativity theory, with its clock time of Einstein, accounts for the periodic motions inside our universe. Clocks are built on periodic motions. There is no arrow of time with ongoing periodic motion, only a hint of the arrow with its decay.

On the other hand, there is no periodic motion in the expansion of space. The expansion-time is an arrow from the past straight into the future.

When he built his relativity theories, Einstein thought our universe was static and so he built no source of expansion of space into his model.

History may well conclude that we are fortunate that Einstein developed his seminal relativity theory before Hubble found our universe was expanding. Had Einstein then needed to also account for expansion, the product might well have been neither.

The author began his development of a new model some 20 years ago in an attempt to account for Newton's gravity without the concept of a field of attraction, but in terms of a "push" instead of a "pull". Success on this first attempt [1-2], led to the development of a new model for the expansion of our universe. Most of that development has been recorded [1-10] and will be finished in a paper in preparation to follow this paper. The goal of this paper, as indicated in its title, is to convince the reader that an acceleration of the expansion rate of our universe is not possible without destroying an increasing time into the future.

Einstein got a hint of this dilemma when he first applied his new general relativity to cosmology [13]. He chose a 3-sphere[1] geometry for his assumed static universe and discovered it was unstable and began to collapse from his built-in "gravitational attraction" of all the mass in the universe. So he added his infamous lambda term to his field equations to cancel the gravitational attraction.

After Hubble showed our universe was expanding, Einstein promptly withdrew his lambda term.. But with measurements of the brightness of exploding SNIa stars, theoreticians have re-introduced his lambda term and dark energy to make their version of relativity fit the data.

However these new additions to the contents of our universe also introduce even more problems of an acceleration of the expansion of the universe. The last two decades will be an embarrassing and costly era in the development of cosmology.

It has been repeatedly shown, but so far to no avail[2], that lambda or dark energy are not needed to fit the SNIa data [3,9,11]. Thus lambda and dark energy do not exist.

What does exist, however, is a much more complicated universe than Einstein conceived. We were warned by the Cleric, E. A. Abbot, via his two dimensional character A-Square in his 1885 classic book, *Flatland* [14], that 3-D occupants had a very small thickness in the fourth dimension "extra-height (EH)".

Indeed we do have an EH-Planck thickness in the radial direction of our three-sphere which is not only the source of gravity but the source of the production of space that expands our universe.

But relativity theory allows only three spatial dimensions, so even though the origin of the sphere is in the center of the spherical cavity, we are told that there is no reality to either the inside or outside of the sphere itself.

Accepting a Planck thickness on the outside of our 3-sphere allows reality to the inside of the 3-sphere.

The process for the production of new space on the outside of the 3-sphere is called Spatial Condensation (SC). But it does not directly produce 3-D.space but Planck-size spatial particles of 4-D space that expand the 4-D ball cavity and thus our 3-D universe on its surface

The all important concept is that the expansion of our 3-D universe is completely free of the gravitational interaction of the masses of our universe. It cannot collapse because of the supporting and expanding 4-D ball. The masses of our 3-D universe continue to locally curve 3-D space and these dimples in the 4-D ball interact according to Newton's equation. By a simple transformation of Newton's equation, the source of that acceleration was shown to be a factor of $\approx 10^{20}$ times greater than the weak component of gravity that we measure here on Earth [8].

This Introduction already reviews some of the fundamental new concepts of the new Spatial Condensation (SC) model. The balance of the paper is organized as follows: Section 1 presents a brief review of the beginning of the SC-process in the older pre-existing epi-universe. Section 2 examines some geometric relations of the new model with a surprising prediction of the expansion rate. Section 3 lists many big bang problems in attempts to account for the expansion with relativity theory. Section 4 presents equations of the closed SC-model which are pertinent to the task of the title to this paper. Section 5 then presents the analysis that shows the conflict with future time for any cosmological model that attempts to predict an accelerated expansion. Section 6 presents the Summary and Conclusions

---

[1] Do a Google on "N-sphere".
[2] At their lectures to: S. Perlmutter (UofM, 11/11/98): M. Turner (UofM,5/17/03); R. Kirshner (WSU, 4/27/05); No interest!

# SECTION 1

## 1.1 The Beginning

The mathematical beginning of the expansion has been presented elsewhere [1,2,9], so a summary in words will serve here. The spatial cells of epi-space are much smaller than even the Planck-size spatial cells of our 3-D space and the number N of spatial dimensions of epi-space is greater than 4.

In a certain region of epi-space, epi-energy has been draining away and can barely support the N spatial dimensions. A spatial condensation process has been trying to get started and then at one point at 3-D time t = 0, that process did go to completion and produced the first cell of 4-D space, a very small hypercube with edges of Planck length (~$10^{-33}$ cm).

Any foreign object in epi-space is a catalytic site for this spatial condensation process. So an exponential production of additional "free" 4-D cells was underway. One new 4-D cell was produced every Planck second on each existing, and exposed, 4-D cell in this region.

A 4-D spatial particle occupies less epi-space than all of the N-D spatial epi-particles that condense to form the 4-D cell. Thus the epi-pressure reduced in this region of spatial condensation, and the inrushing epi-cells drove the "free" 4-D cells into a 4-D spatial ball. This isolated most of the 4-D cells inside the 4-D ball and now the surface of the 4-D ball was the only foreign object in the epi-universe for continued spatial condensation, but at a much reduced rate.

With t = 0 at production of the first 4-D cell, our 3-D universe was produced at the finite time of ~$10^{-33}$ seconds.

Note the three fundamental hypotheses at the very beginning of this new cosmological model: (1) the pre-existence of a higher-dimensional (N > 4) epi-universe, that provided the spatial building blocks for the production of our: (2) 3-sphere closed universe that consist of (3) an expanding 4-D ball with our 3-D universe on its surface.

Obviously we cannot enter or perform experiments in these higher dimensional spaces. So their existence must be supported by existing measurements that cannot be explained without the acceptance of these three new hypotheses.

The details of the formation of the 4-D spatial ball are very important; for example: pre-collisions of smaller with larger 4-D balls in the presence of the beginnings of matter mass. Much of the later large-scale structure of our present universe was fixed in that first ~$10^{-33}$ second. Some details will be added shortly but now the consequences of the production of 4-D space must be made clear.

## 1.2 Planck Natural Units[3]

In his seminal paper of 1899 where he presented his Planck constant h [15], he also recognized, in an appendix, that it together with the gravitational constant G, and the speed of light c, defined a set of natural units for length, time and mass. In my cgs units, they are $l_p = 1.616 \times 10^{-33}$ cm, $t_p = 5.391 \times 10^{-44}$ s and $m_p = 2.177 \times 10^{-5}$ g, where, $\hbar = h/2\pi$.

As will be shown, these units were very important to the development of the SC-model. Both $l_p$ and $t_p$ are much smaller than can be measured. Even so, nature can not know what values we have selected for our units of measurement, so it is often useful to work in dimensionless numbers such as $N_R = R/l_p$, $N_t = t/t_p$ and $N_M = M/m_p$.

## 1.3 Observables

Certain quantities are known within limits, about our 3-D universe after much study and astronomical measurements such as WMAP [16]. Among these quantities are its age: $t_0 = 13.7 \pm 0.2$ Gy, Hubble $H_0 = 71^{+4}_{-3}$ km s$^{-1}$ Mpc$^{-1}$ and its size ≈ Hubble length $R_0 = c/H_0 \approx 4225$ Mpc. SC-values of $t_0$ = 13.5, $H_0$ = 68.61 and $R_0$ = 4388 are in good agreement.

---

[3] Do a Google on Planck natural units

# SECTION 2
## 2.1 Geometric Considerations

The closed geometry of a 3-sphere was selected for the geometry of our 3-D universe. The presentation of the mathematical space-time relations will be made in Section 4, but a preliminary examination of the geometry is now made.

The volume of a 4.D ball is $V_4 = 1/2\pi^2 R^4$ and the volume of its 3-D surface is $V_3 = 2\pi^2 R^3$. The volume of a 4-D spatial particle is $l_p^4$ and of a 3-D spatial particle is, $l_p^3$. So to work with non-dimensional numbers, the number of spatial particles in the 4-ball is, $N_4 = V_4/l_p^4$ and the number in its surface is $N_3 = V_3/l_p^3$.

Continuing with the beginning hypothesis that one new 4-D spatial cell is produced every $t_p$ second on each exposed 4-D cell, and the number of 4-D cells exposed on the surface is $N_3$, then we can write that the present production rate of 4-D cells is,

$$\dot{V}_{40} = N_{30} l_p^4 / t_p \qquad (1)$$

Now pause a moment to reflect on this equation. It is not the type of equation that normally occurs in mathematics.
This rate of increase of the 4-D volume is composed entirely of Planck natural units. It is as if both space and time have been quantized to cellular spaces and discrete time.

In Newton's calculus, which we will use later, $\dot{V} = dV/dt = \lim_{\Delta t \to 0} \Delta V/\Delta t$.
In Eq. (1), the limit stops at Planck, $t_p$

Inserting values in Eq. (1), with SC-$R_0$ = $1.354 \times 10^{28}$ cm, $N_{30} = 1.161 \times 10^{183}$ and $t_p$,
$\dot{V}_{40} = 1.470 \times 10^{96}$ cm$^4$ s$^{-1}$ = $5.15 \times 10^5$ Mpc$^4$/yr, which appears reasonable.

Taking ordinary derivatives,
$\dot{V}_3 = 6\pi^2 R^2 \dot{R}$ and $\dot{V}_4 = 2\pi^2 R^3 \dot{R}$, so
$\dot{V}_3/\dot{V}_4 = 3/R$ or $\dot{V}_{30} = 3.272 \times 10^{68}$ cm$^3$ s$^{-1}$ = 325 Mpc$^3$/yr.

Borrowing mathematical (math) time from Section 4 for the abscissa, the spatial production rates $\dot{V}_3$ and $\dot{V}_4$ are plotted in Fig. 1, divided by their values at $t_0 = 13.5 Gy$,

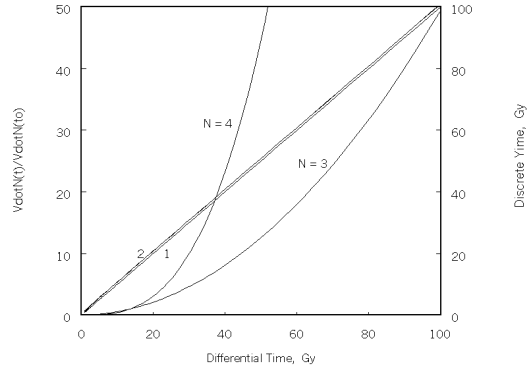

Fig 1. Vdot3($t_0$)=3.272x10$^{68}$cm$^3$/s $t_0$=13.5Gy; Vdot4($t_0$)=1.477x10$^{96}$ cm$^4$/s.
Discrete times:
$t_2$ = R/c, $t_3$=3$V_3$/Vdot3, $t_4$=4$V_4$/Vdot4, all on curve 2. Eq. (8) time (curve1) = Math.-time

Remember the processing hypothesis of spatial condensation is: *one new 4-D cell per Planck time on each exposed 4-D cell*. That is how SC started. Next, imagine a new slightly longer radius R to each exposed 4-D cell on the surface of the ball. That is continued expansion.

Since, $c = l_p/t_p$, $R/c = N_R t_p$, which is a period of discrete time; call it $t_2$ time. Curve 2 in Fig. 1 is $t_2$ discrete time.

From Eq. (3), $t = \sqrt{\kappa/G\rho}$; this is math time on diagonal curve1 of Fig. 1. Since they equal R/c, expressions $t_3 = 3V_3/Vdot3$ and $t_4 = 4V_4/Vdot4$ are also discrete times plotted on curve 2.

The author found that if each $t_2$ $t_3$, $t_4$ is multiplied by $\rho/\rho_2$, Eq. (9), then all $t_2$, $t_3$, t4 curves also fell on math curve 1. Present value of $\rho/\rho_2 = tH = 0.9473$.

The dimensionless ratio of densities with range $1/2 \leq \rho/\rho_2 \leq 1$, accounts for the changing contents of our universe, but that is a weak argument for its arbitrary multiplication. We dig deeper and continue with the expansion of Eq. (9)

Replace H by $\dot{R}/R$ and divide both sides by c to get $R/c = (t\dot{R}/c)\rho/\rho_2$. Expand, $c = l_p/t_p$. On the left we get, $(N_R/t_p)(\rho/\rho_2)$ and on the right $t(dR/l_p)/(dt/t_p) = t(dN_R/dN_t)$. But $dN_R/dN_t = 1$ is a statement of the processing hypothesis, that there is a unit increase in the number of Planck lengths in the radius R per unit change of Planck time. Thus we get a theoretical justification for the multiplication of discrete time by $\rho/\rho_2$ to get math time,

$$t = (N_R/t_p)(\rho/\rho_2) \qquad (2).$$

The author will leave any continued development of these concepts, and any consequences for quantum theory, to other scientists.

## SECTION 3
### 3.1 List of Big Bang Problems
Most text books on cosmology list the fundamental problems of the Big Bang theory. Many authors select their most fundamental problem [17-19].

### 3.1.1 Expansion Problem
The Friedmann equation solution to Einstein's field equations is not in terms of time but in terms of the Hubble parameter H,

$\dot{R}^2 + kc^2 = (8\pi G/3)R^2\rho$ or $H^2 + kc^2/R^2 = (8\pi G/3)\rho$

Immediately, this equation presents problems. It says that the radius R increases because its contents, the density, increase. In the vernacular, the universe is expanding itself by its own bootstraps. That false concept will not work. A reasonable production of space must be found. John Peacock in his book *Cosmological Physics* [17, p 324] selected this Expansion Problem as the most profound of the BB-theory. He also discussed other problems.

The reader has already read the SC-solution to this problem.

### 3.1.2 Horizon Problem
In the early BB-universe, that contains all of the present mass of our universe, a particle horizon exists because photons have not had time to visit all of the early space and provide a means to account for all of the large-scale uniformity of the 3-D space we see today.

This problem does not exist in the SC-model that starts at t = 0 from a single 4-D spatial particle and expands violently before our 3-D universe was created with the beginnings of mass.

### 3.1.3 Flatness Problem
Present BB-cosmologists prefer the flat, infinite universe of the k=0 Friedman universe. One reason is that they can use simpler Euclidian geometry. The requirement, $\Omega = 1$, makes the model unstable. No geometry could be more different from the SC closed 3-sphere geometry. Also how can space expand in an infinite no-boundary universe?

### 3.1.4 Anti-matter Problem
This BB-problem arises because equal mass of matter and anti-matter (none permanent at present) was expected to be produced in the early universe. This problem is not addressed in the SC-model.

### 3.1.5 The Structure Problem
Our 3-Duniverse is not uniform on the 1000 Mpc scale. But there is nothing built into the BB-model to account for that non-uniformity. SC-expansion forces would erase any early spatial discontinuities, but the SC fast-growing globs of dark mass form early black holes and do seed the galaxies.

### 3.1.6 Vacuum Energy Problem

Quantum theory predicts that if vacuum zero-point energy is cut off at the Planck scale, the present ratio of vacuum energy to mass energy of our 3-D universe is the impossible enormous ratio of $\sim 10^{123}$.

It turns out that the SC-model predicts exactly the same ratio $2 \times 10^{123}$ but with the major difference that vacuum energy is a new form of energy that does not carry the attribute of mass.

### 3.1.7 Fundamental Problems

These problems are not only for BB-theory but for all of physics. Relations are well known between the various forms of energy, momentum and mass. But at a deeper depth of understanding, the motions in our universe that account for energy, momentum and mass are completely lacking. Spatial condensation should provide that missing understanding.[5]

### 3.1.8 Inflation Problems

This problem was added to the BB-model in an attempt to solve some of the problems above such as 3.1.2. Very shortly after the big bang, it was postulated that there was an even much greater bang that stretched the boundary of our no-boundary $\Omega = 1$ universe so that photons had ample tine to visit all of our early 3-D space—Energy source? It seems, just add a large capital V for potential energy at the end of a Lagrangian.

### 3.1.9 Source of Energy

The eminent cosmologist Edward R. Harrison [20] was adamant that there is no conservation of energy in BB-theory. Indeed, his last paper [21] showed that, in principle, useful energy could be generated by tethering two galaxies, even if there is no accounting for a source of energy in BB theory.

In the SC-model, the source of energy for our 3-D universe, is obtained by adding the interacting epi-universe to the total system.

This is proven every day when one lifts an object from the floor to the table. The work against gravity is stored back into the epi-universe, from whence it came as it fell to the floor.

### 3.1.10 Future Problem

This is the Title-Problem to be addressed in much more detail in Section 5.

The expansion scale factor $R/R_0$ increases with expansion time into the future. But the expansion redshift $z = 1/(R/R_0)-1$ deceases with increasing $R/R_0$ into the future. As shown in Fig. 2, the entire future of our universe is contained between $z = 0$ and $z = -1$.

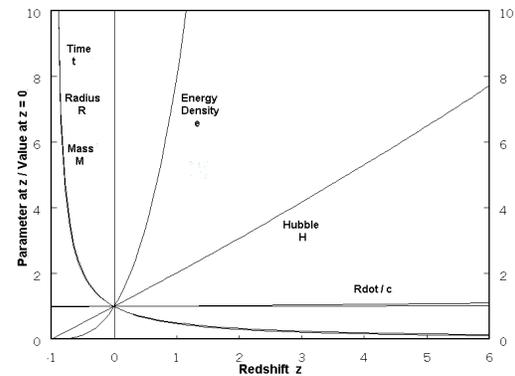

Fig. 2 SC-parameters are well behaved into the future $-1 \leq z \leq 0$ with $z = 1/(R/R_0) - 1$ Parameter values at z were divided by their value at z=0.

The future BB-problems will be discussed in Section 5.

## SECTION 4
### 4.1 SC-Decelerated Expansion

The SC-law of expansion is contained in the simple dimensionless expression,

$$G\rho t^2 = \kappa = 3/32\pi \qquad (3)$$

Where $\rho$ is the total slowly changing density of our spatially 3-D universe, G is the gravitational constant and finally, we have a model where time t, s, is the fundamental parameter and not the Hubble parameter H, $s^{-1}$ of BB-theory.

Note that $t = \sqrt{(\kappa/G)/\rho}$, so that it is clear that the sum of all densities must decrease with increasing time into the future.

Other pertinent equations are.

$$\rho_2 = \rho_{r0}(R_0/R)^4 + \rho_{m0}(R_0/R)^3 + \rho_{x0}(R_0/R)^2 \quad (4)$$

$$\rho' = \partial\rho/\partial R = -2\rho_2/R, \quad (5)$$

$$\rho_2 = 2\rho_r + (3/2)\rho_m + \rho_x. \quad (6)$$

$$H(z) = H_0 \left(\frac{\rho_{20}}{\rho_0^{3/2}}\right)\left(\frac{\rho^{3/2}(z)}{\rho_2(z)}\right) \quad (7)$$

$$t(z) = t_0 \rho_0^{1/2} x \left[\rho_{r0}(1+z)^4 + \rho_{m0}(1+z)^3 + \rho_{x0}(1+z)^2\right]^{-1/2} \quad (8)$$

$$tH = \rho/\rho_2, \quad [1/2 \le tH \le 1] \quad (9)$$

$\rho_{r0} = 9.40 \times 10^{-34}$ and $\rho_{m0} = 2.72 \times 10^{-31}$
$\rho_{x0} = \kappa/G t_0^2 - \rho_{r0} - \rho_{m0}$ and $t_0 = 13.5\, Gy$

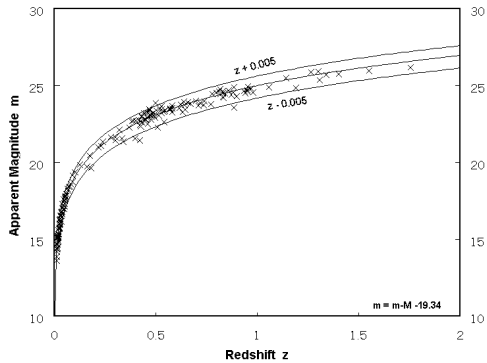

Fig. 3 Published 185 "gold and silver" SNIa data and SC-curve. No added lambda or dark energy were needed to fit the data.

Thus as shown in Fig. 3, the SC-model does indeed fit the astronomical measurements of radiation from exploding SNIa stars [22] without any lambda or dark energy.

The center curve is the SC-prediction of apparent magnitude m. The two outside curves is the same but with built in errors of $z \pm 0.005$ mag.

Besides the SC-concepts presented here, there is yet another very important concept, very different from relativity theory that was needed to produce the good fit of Fig. 3. That new concept, and its support, will be presented in the next paper in preparation.

However that concept was not needed for the good fit in Fig. 4

Luminosity measurements were not needed for measurement of the Hubble parameter H of eight passively separating galaxies [23, 24] as shown in Fig.4. The SC-predicted curve of $H/H_o$ versus scale factor $R/R_o$ fits the data well.

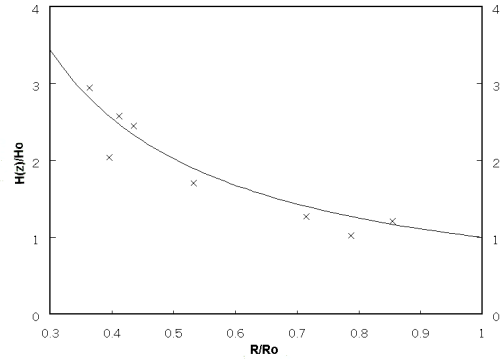

Fig. 4 For $H(z)/H_0$ obtained from passively separating galaxies, the eight reported values had a low chi value of 0.043.

## SECTION 5
**5.1 Big Bang Future Problem**
The scale factor $R/R_0$ sets the size of our universe relative to its present size, radius $R_0 \approx c/H_0 \approx 4371\, Mpc$. Astronomers also use redshift z of arriving radiation to set the size of our universe when that radiation was emitted. The relation between these two measures is.

$$R_0/R = (1+z)\ or\ z = 1/(1+z) - 1 \quad (10)$$

The present is $R/R_0=1$ or $z = 0$. The future is $1 \leq R/R_0 \leq \infty$ or $-1 \leq z \leq 0$. Thus any size of our universe in the future is represented by a negative number between $-1 \leq z \leq 0$. Of course, we do not receive radiation from the future, but the equations for derived physical parameters should extend into the future as they do for SC-H and SC-time t in Fig. 2.

Rarely does a cosmologist, using relativity theory, extend a physical parameter into the future. Instead they cut off their parameter at z = 0. Incorrect models generally predict non-physical behavior into the future.

In particular, any model that predicts an accelerated expansion rate, will also predict an unphysical behavior of time into the future. The reason is stated for the SC-model by Eq. (9) since $\rho/\rho_2 \approx 1$. If time is to increase freely into the future, then the Hubble parameter H must go to zero into the future. The same limit applies to relativity theory {17, p 85}.

Using the current GR-standard $\Lambda CDM$ model, $(\Omega_m = 0.28, \Omega_\Lambda = 0.72)$, the predicted unphysical behavior of BB-H and BB-time t are shown in Fig. 5.

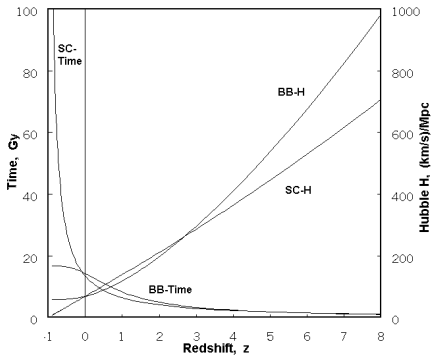

Fig. 5 SC-curves (see Fig.2) are well behaved into the future; whereas lambda of ($\Omega_m =0.28$, $\Omega_\Lambda = 0.72$) of the BB-model drives BB-H and BB-t flat, into the future.

The acceleration due to lambda, $\Omega_\Lambda = 0.72$, drives BB-H flat into the future and, in turn, that drives BB-time flat and unphysical. As in Fig. 2, the SC-H and SC-time in Fig. 5 are well behaved in the future.

## SECTION 6
### 6.1 Summary and Conclusions

Section 1 reviewed the background and basics of the new SC-model for the expansion of the universe. In Section 2 an analysis of geometric considerations, even before the new mathematical model was presented, disclosed an alternative quantum foundation of cellular spaces and discrete time and the transformation between the two. Section 3 listed the many problems of the Big Bang model in its use to account for the expansion of our universe. Included were discussions of how the SC-model solved or avoided these problems. Section 4 presented the fundamental equations of the new SC-closed model with no free parameters which serve as the basis for demonstrating that acceleration of the expansion rate is not possible in an acceptable model of our universe. Section 5 did show that the Big Bang model with its standard $\Lambda CDM$ contents, produces in the future z<0, an unphysical Hubble parameter and unphysical expansion time.

In contrast, the closed SC-model with its decelerated expansion has no problem with its predictions into the future.

Probably, the most important conclusion of this paper is that the Epi-universe must be included in the total cosmological system, not only to provide a source of space for the expansion of our universe, but also to re-establish the important conservation of energy.

### ACKNOWLEDGMENTS

The author thanks his good friend .Emeritus Professor Robert A Piccirelli, for extensive discussions of the new physical concepts.